\documentclass[preprint,showpacs,showkeys,preprintnumbers,amsmath,amssymb]{revtex4}
\usepackage{epsf}
\usepackage{color}
\usepackage{dcolumn}
\usepackage{bm}
\everymath{\displaystyle}
\begin{document}

\title{Connection Between Wave Functions in the Dirac and
Foldy-Wouthuysen Representations}

\author{Alexander J. Silenko}
\email{silenko@inp.minsk.by}

\affiliation{Institute of Nuclear Problems, Belarusian State
University, Minsk 220080, Belarus}

\date{\today}

\begin {abstract}
The connection between wave functions in the Dirac and
Foldy-Wouthuysen representations is found. When the Foldy-Wouthuysen
transformation is exact, upper spinors in two representations differ
only by constant factors, and lower spinors in the Foldy-Wouthuysen
representation are equal to zero.
\end{abstract}

\maketitle

\section {Introduction}

Determination of connection between wave functions in the Dirac and
Foldy-Wouthuysen (FW) representations is a very important problem.
The Hamiltonian for relativistic particles in the FW representation
contains a square root of operators (see Refs. \cite{FW,JMP}).
Therefore, the Dirac representation is usually more convenient than
FW one for finding wave eigenfunctions and eigenvalues of the
Hamilton operator. However, the form of operators is much simpler
just in the FW representation. For example, in this representation
$\bm r$ is the coordinate operator, $\bm p=-i\nabla$ is the momentum
operator, $\bm l=\bm r\times\bm p$ is the orbital angular momentum
operator, the matrix $\bm\Sigma/2$ is the spin momentum operator
(see Ref. \cite{FW}). The polarization operator in the FW
representation is the matrix $\bm\Pi$ \cite{FG}. In the Dirac
representation, all corresponding operators are defined by
cumbersome expressions \cite{FW}. For particles interacting with
external fields, these operators also depend on the external field
parameters.

The FW representation is very useful to derive semiclassical
equations of particle and spin motion \cite{JMP}.

The determination of connection between wave functions in the Dirac
and FW representations makes it possible to calculate wave functions
only in one of these representations. In the present work, the
connection between wave functions is derived when the FW
representation is exact.

   Throughout the work we use the system of units $\hbar=c=1$
   and generally accepted designations of Dirac matrices (see \cite{FW}).

\section {The exact Foldy-Wouthuysen transformation}

   In the general case, the transformation to a new representation described by the wave
function $\Psi'$ is performed with the unitary operator $U$:
$$\Psi'=U\Psi,$$
where $\Psi$ is the wave function (bispinor) in the Dirac
representation. The Hamilton operator in the new representation
takes the form  \cite{FW,Gol} $$ {\cal H}'=U{\cal H}U^{-1}-iU\frac{
\partial U^{-1}}{\partial t}. $$

   The Hamiltonian can be split into
operators commuting and noncommuting with the operator $\beta$:
\begin{equation} {\cal H}=\beta m+{\cal E}+{\cal
O},~~~\beta{\cal E}={\cal E}\beta, ~~~\beta{\cal O}=-{\cal
O}\beta. \label{eq1} \end{equation}

The FW transformation is exact if the external field is stationary
and the operators ${\cal E}$ and ${\cal O}$ commute \cite{JMP}:
\begin{equation}
[{\cal E},{\cal O}]=0. \label{eq2}
\end{equation}

Eq. (2) is a sufficient but not necessary condition of the exact
transformation. If this condition is satisfied, the Hamilton
operator in the FW representation is exactly defined by \cite{JMP}
\begin{equation}
{\cal H}_{FW}=\beta \epsilon+{\cal E},~~~ \epsilon=\sqrt{m^2+{\cal
O}^2}. \label{eq3}
\end{equation}

   In this case, the transformation operator $U$ is described by
\begin{equation}
U^{\pm}=\frac{\epsilon+m\pm\beta{\cal
O}}{\sqrt{2\epsilon(\epsilon+m)}}, \label{eq4} \end{equation}
 where $U^+\equiv U,~U^-\equiv U^{-1}$.

It follows from formulae (2),(3) that the operators $\beta{\cal
O}^2$ and ${\cal E}$ commute with the Hamiltonian ${\cal H}_{FW}$.
Therefore, wave eigenfunctions in the FW representation are also
eigenfunctions of the operators $\beta{\cal O}^2$ and ${\cal E}$.
This circumstance simplifies considerably the determination of wave
eigenfunctions.

\section {The Foldy-Wouthuysen transformation of several operators}

   Due to form (4) of operator $U$, some operators remain unchanged after the FW
transformation for a stationary external field:
\begin{equation} {\cal E}_{FW}={\cal E}, ~~~\left(i\frac{\partial}
{\partial t}\right)_{FW}=i\frac{\partial}{\partial t},
~~~\epsilon_{FW}= \epsilon. \label{eq5} \end{equation}

Eqs. (1)--(3) show that in the Dirac representation the operators
${\cal E}$ and $\epsilon$ commute with the Hamiltonian. Therefore,
they have the eigenvalues, ${\cal E}_0$ and $\epsilon_0$:
\begin{equation} {\cal E}\Psi={\cal E}_0\Psi, ~~~\epsilon\Psi=
\epsilon_0\Psi. \label{eq6} \end{equation}

The operators ${\cal E}$ and $\epsilon$ also commute with the
Hamiltonian in the FW representation, ${\cal H}_{FW}$. The
eigenvalues of these operators are defined by
\begin{equation} {\cal E}\Psi_{FW}={\cal E}_0\Psi_{FW}, ~~~\epsilon\Psi_{FW}=
\epsilon_0\Psi_{FW}. \label{eq7} \end{equation}

   Certainly, the eigenvalues in Eqs. (6) and (7) coincide.
For the particle energy operator, similar equations take the form
\begin{equation}
i\frac{\partial}{\partial t}\Psi={\cal H}\Psi=E\Psi, ~~~
i\frac{\partial}{\partial t}\Psi_{FW}={\cal
H}_{FW}\Psi_{FW}=E\Psi_{FW}. \label{eq8} \end{equation}

\section {Connection between wave functions}

   Let us analyze a connection between Dirac and FW wave functions
   when
the FW transformation is exact. Eq. (1) leads to the formula
$$ i\beta\frac{\partial}{\partial t}\Psi=(m+\beta{\cal E}+\beta{\cal O})\Psi.
$$

   Therefore, the connection between initial and final wave functions takes the
form
\begin{equation}
\Psi_{FW}=U\Psi=\frac{\epsilon+i\beta\frac{\partial}{\partial
t}-\beta {\cal E}}{\sqrt{2\epsilon(\epsilon+m)}}\Psi. \label{eq9}
\end{equation}

   Formulae (6),(8),(9) lead to the relation
\begin{equation}
\Psi_{FW}=\frac{\epsilon_0+\beta(E-{\cal
E}_0)}{\sqrt{2\epsilon_0(\epsilon_0 +m)}}\Psi. \label{eq10}
\end{equation}

   This relation shows wave functions in the Dirac $\Psi=
\left(\begin{array}{c}\phi \\ \chi\end{array}\right)$ and FW
representations differ only by constant factors:
$$\Psi_{FW}=\left(\begin{array}{c} c_1\phi \\ c_2\chi\end{array}\right)=
\left(\begin{array}{cc}c_1&0 \\ 0&c_2\end{array}\right)
\left(\begin{array}{c}\phi \\ \chi\end{array}\right). $$

   Formulae (3),(8) result in
\begin{equation}
(\beta\epsilon+{\cal E})\Psi_{FW}=E\Psi_{FW}. \label{eq11}
\end{equation}

   In accordance with Eqs. (7),(11), either
$E=\epsilon_0+{\cal E}_0$ or $E=-\epsilon_0+{\cal E}_0$. Since the
total energy of particles is positive, first of these relations is
right. As a result, Eq. (10) takes the form
\begin{equation}
\Psi_{FW}=\frac{(1+\beta)\epsilon_0}{\sqrt{2\epsilon_0(\epsilon_0+m)}}\Psi=
\sqrt{\frac{2\epsilon_0}{\epsilon_0+m}}\left(\begin{array}{c}\phi \\
0  \end{array}\right). \label{eq12} \end{equation}

   It follows from Eq. (12) that upper spinors in the Dirac and FW
representations differ only by constant factors and lower spinors
in the FW representation are equal to zero. Wave functions are
normalized to unit. Therefore, we can renormalize the wave
functions instead of calculation of the value $\epsilon_0$. As a
result, the wave function in the FW representation can also be
given by
\begin{equation}
\Psi_{FW}=\frac{1}{\int{\phi^{\dag}\phi dV}}\left(\begin{array}{c}\phi \\
0 \end{array}\right). \label{eq13} \end{equation}

In the considered case, a solution of wave equations in the FW
representation is quite possible. Therefore, the connection between
wave functions in two representations makes it possible to deduce FW
wave eigenfunctions from Dirac wave eigenfunctions or on the
contrary. In the Dirac representation, the lower spinor can be
expressed in terms of the upper one.

\section {Example: particle in a uniform magnetic field}

 The FW transformation is exact for particles with an anomalous magnetic
moment (AMM) moving in the plane orthogonal to a static uniform
magnetic field \cite{JMP}. In this case, the operator
$p_z\!=\!-i(\partial /\partial z)$ commutes with the Hamilton
operator and has eigenvalues $P_z\!=\!{\rm const}$. Therefore, the
consideration of the particular case $P_z\!=\!0$ is quite reasonable
\cite{JMP}.

   The Hamilton operator in the Dirac representation satisfies Eq. (1) where
\begin{equation}      \begin{array}{c}
{\cal E}=-\mu'\bm {\Pi}\cdot \bm{H},~~~ {\cal
O}=\bm{\alpha}\cdot\bm{\pi}, ~~~\bm{\pi}=\bm{p}-e\bm{ A},
\end{array} \label{eq14} \end{equation} $\mu'$ is AMM, $\bm{ A}$ is the vector
potential, and $\bm{H}$ is the magnetic field strength.

   In this case, the Hamilton operator in the FW representation is equal to
\cite{JETP}
\begin{equation}
{\cal H}_{FW}=\beta\sqrt{\bm\pi^{\,2}+m^2-e\bm {\Sigma} \cdot\bm
H}-\mu'\bm\Pi\cdot \bm H, \label{eq15} \end{equation} where
$\pi_z\Psi_{FW}=P_z\Psi_{FW}=0$ and
$$\epsilon=\sqrt{\bm\pi^{\,2}+m^2-e\bm {\Sigma} \cdot\bm H}
=\sqrt{\bm\pi^{\,2}_{\bot}+m^2-e\bm {\Sigma} \cdot\bm H}. $$

   The eigenvalues of the operators are defined by \cite{Ts,TBZ}
\begin{equation}      \begin{array}{c}
\epsilon_0=\sqrt{m^2+(2n+1)|e|H-\lambda eH},~~~
{\cal E}_0=-\lambda \mu'H, \\ n=0,1,2,\dots,~~~
\lambda=\pm 1.
\end{array}\label{eq16} \end{equation}

   Therefore, the connection between wave functions is given by Eq.
   (12), where $\phi$ is the upper spinor in
the Dirac representation and $\epsilon_0$ is defined by Eq. (16).

Hamiltonian (15) commutes with the operators $\bm\pi^{\,2}_{\bot}$
and $\Pi_z$. Therefore, the wave eigenfunction $\Psi_{FW}$ is also
an eigenfunction of these operators and can be given by
$$\Psi_{FW}=\psi\zeta,$$ where $\psi$ is a coordinate wave function
and $\zeta$ is an eigenfunction of operator $\Pi_z$:
$$\Pi_z\zeta=\lambda\zeta, ~~~ \lambda=\pm1.$$ Since the lower spinors are zero,
$\zeta^+=\left(\begin{array}{c} 1 \\ 0 \\
0 \\ 0\end{array}\right)$ when $\lambda=1$, and
$\zeta^-=\left(\begin{array}{c} 0 \\ 1
 \\0 \\ 0\end{array}\right)$ when $\lambda=-1$.

 We should take into account $z$-components of spin and orbital angular momentum
 have definite values. When $z$-component of total angular
 momentum equals $M$, the wave eigenfunctions take the form
\begin{equation}\begin{array}{c}
\Psi_{FW}^+=\frac{\exp{[i(M-1/2)\varphi}]}{\sqrt{2\pi}}R_{|M-1/2|}(\rho)\zeta^+,\\
\Psi_{FW}^-=\frac{\exp{[i(M+1/2)\varphi]}}{\sqrt{2\pi}}R_{|M+1/2|}(\rho)\zeta^-,
\end{array} \label{eq17} \end{equation}
 where $R_{|M\pm1/2|}(\rho)$ are the well-known radial eigenfunctions of operator
$\bm\pi^{\,2}_{\bot}$ and the signs "$+$" and "$-$" mean positive
and negative projections of spin, respectively.

For the particular case $P_z=0$, the wave eigenfunctions of Dirac
particles with AMM in the uniform magnetic field derived in Ref.
\cite{TBZ} can be obtained from Eq. (17).

\section {Summary}

   In this work, the connection between wave functions in the Dirac and
Foldy-Wouthuysen representations has been found. When the FW
transformation is exact, upper spinors in two representations differ
only by constant factors, and lower spinors in the FW representation
are zero. In the FW representation, relativistic formulae for
operators are the simplest \cite{FW,JMP}. Therefore, it is possible
to use the wave eigenfunctions in the Dirac representation for
calculating the mean values of operators in the FW representation.
In other cases, the Dirac wave eigenfunctions can be deduced from
the FW ones.

\section {Acknowledgements}

I would like to thank Prof. V.A. Borisov for helpful discussion and
to acknowledge a financial support by the BRFFR.

\end{document}